\documentclass[twocolumn]{aa}
\usepackage{graphicx}
\usepackage{txfonts}
\usepackage{natbib}
\bibpunct{(}{)}{;}{a}{}{,}
\begin{document}
   \title{The Gamma-ray burst 050904 : evidence for a termination shock ?}

   \author{B. Gendre  \inst{1}, A. Galli \inst{1}$^,$~\inst{2}$^,$~\inst{3}, A. Corsi \inst{1}$^,$~\inst{2}$^,$~\inst{4}, A. Klotz \inst{5}$^,$~\inst{6},  L. Piro \inst{1}, G. Stratta \inst{7}, M. Bo\"er \inst{6}, \and Y. Damerdji \inst{6}$^,$~\inst{5}}

   \offprints{B. Gendre}

   \institute{IASF-Roma/INAF, via fosso del cavaliere 100, 00133 Roma, Italy \\
		\email{bruce.gendre@iasf-roma.inaf.it, alessandra.corsi@iasf-roma.inaf.it, galli@iasf-roma.inaf.it, piro@iasf-roma.inaf.it}
	\and
          Universit\'a degli Studi di Roma "La Sapienza", Piazzale A. Moro 5, 00185, Roma, Italy
	\and
          INFN - Sezione di Trieste c/o Dipartimento di Fisica $-$ Universit\'a di Trieste  Via Valerio 2, 34127 Trieste, Italy
        \and
          INFN - Sezione di Roma c/o Dipartimento di Fisica $-$ Universit\'a degli Studi di Roma "La Sapienza" Piazzale A. Moro 5, 00185 Roma, Italy
        \and
		CESR, 9 avenue du Colonel Roche, 31400 Toulouse, France\\
		\email{klotz@cesr.fr}
         \and
             Observatoire de Haute-Provence, 04870 St Michel l'Observatoire, France\\
                \email{michel.boer@oamp.fr}
	\and
 	LATT, Observatoire Midi-Pyr\'en\'ees, 14 avenue Edouard Belin, 31400 Toulouse, France\\
              \email{gstratta@ast.obs-mip.fr}
             }

   \date{Received ---; accepted ---}

  \abstract
   {}
   {We analyze optical and X-ray observations of GRB 050904 obtained with TAROT and  SWIFT.}
   {We perform temporal and spectral analysis of the X-ray and optical data.}
   {We find significant absorption in the early phase of the X-ray light curve, with some evidence (3$\sigma$ level) of variability. We interpret this as a progressive photo-ionization. We investigate the environment of the burst and constrain its density profile. We find that the overall behavior of the afterglow is compatible with a fireball expanding in a wind environment during the first 2000 seconds after the burst (observer frame). On the other hand, the late (after 0.5 days, observer frame) afterglow is consistent with an interstellar medium, suggesting the possible presence of a termination shock. We estimate the termination shock position to be $R_t \sim 1.8 \times 10^{-2}$ pc, and the wind density parameter to be $A_* \sim 1.8$. We try to explain the simultaneous flares observed in optical and X-ray bands in light of different models : delayed external shock from a thick shell, inverse Compton emission from reverse shock, inverse Compton emission from late internal shocks or a very long internal shock activity. Among these models, those based on a single emission mechanism, are unable to account for the broad-band observations. Models invoking late internal shocks, with the inclusion of IC emission, or a properly tuned very long internal shock activity, offer possible explanations.}
   {}


\authorrunning{Gendre et al.}

\titlerunning{Multi wavelength observation of GRB 050904}

   \maketitle
%

\section{Introduction}

Long Gamma-Ray Bursts (GRBs) are cosmological explosions \citep{met97}, thought to be produced by a massive star \citep{mac99}. Their extreme luminosity \citep[up to L$_{iso} \sim 10^{53}$ erg s$^{-1}$,][]{pir05} makes them observable up to very large distance (z$>$ 5). Their long lasting afterglow allows one to study them for days after the trigger \citep[for a recent review see][]{mes06}. 

With the launch of the SWIFT satellite \citep{geh05}, it has been shown that strong flaring activity in the early X-ray afterglow originally found in a few cases by BeppoSAX \citep[e.g.][]{piro05} is a common phenomenon \citep{obr06}. The origin of these X-ray flares is not clear, and several models have been proposed: late internal shocks \citep{Fan05, bur05a}, refreshed energy injection due to long lasting activity of the GRB progenitor \citep{zha05}, rising of an extra spectral component such as inverse Compton \citep{Kob05}, late internal and external shocks \citep{wu_05}, delayed external shock emission in a thick shell fireball \citep{piro05, gal05}. While some of these flares show spectral evolution and may be linked to the prompt emission \citep{bur05a}, others look very similar to the late afterglow emission and can be linked to its onset \citep{piro05, gal05}. Nevertheless, no multi-wavelength observations are usually available to discriminate these hypothesis, and several models can fit the data \citep[e.g \object{GRB 050406},][]{wu_05}. The time dilation, due to cosmological effects, mostly evident for high-z bursts, allows to study more easily the early afterglow and the prompt-to afterglow transition. Bursts located at high distances are thus very interesting to address these issues.

Another interesting aspect of the afterglow studies is the constraints one can put on the absorption around the burst \citep{str04, cam05}. The absorption in the X-ray band is primarily determined by metals. Using the low energy part of the X-ray spectrum, one can thus constrain the metal content of the burst surrounding medium. This is a key issue for high-z bursts, as it allows to study the metal enrichment of the Universe at early epochs.

A final aspect of the afterglow evolution is the problem of the surrounding medium of the burst. Since long GRBs are related with massive stars, one should expect a wind environment around them \citep{che04}. On the other hand, several GRBs exhibit afterglows consistent with an expansion in a constant density medium (hereafter InterStellar Medium, or ISM) rather than a wind environment \citep{pan01}. Several authors \citep[e.g.][]{ram01} proposed that the expanding wind arising from the star can be stopped by a dense surrounding interstellar medium. The interface between these two media is called the termination shock. In this scenario, we would expect the fireball expansion to be consistent with a wind only in the early phase of the afterglow, and consistent with an ISM later \citep{che04, ram01}. However, such a transition was never observed so far. Here we present a multi-wavelength analysis of the \object{GRB 050904} afterglow and show its consistency with the occurrence of a termination shock.

\object{GRB 050904} was at z$=6.29$ and presented strong and long lasting multi-wavelength flaring activities \citep{kaw05, wat05, boe05}. We summarize previous observations in Sec. \ref{sec_present} and describe the data analysis in Sec. \ref{sec_ana}. We report in particular a more detailed analysis of the X-ray and TAROT data. We investigate the absorption evolution around this burst in Sec. \ref{sec_nh}, and the nature of the progenitor together with the medium density profile in Sec. \ref{sec_discu}. We discuss the origin of the observed multi-wavelength flare in the framework of a delayed external shock emission due to a thick shell fireball in Sec. \ref{sec_galli}. Other models proposed in the literature are discussed in Sec. \ref{sec_corsi}. Conclusions are given in Sec. \ref{sec_conclu}.


\section{\object{GRB 050904}}
\label{sec_present}

GRB 050904 was triggered by the BAT instrument \citep{bar05} on board the SWIFT satellite \citep{geh05} on September 4th, 2005, at 01:51:44 UT \citep{cum05}. This was a bright burst, with a fluence of $5.4 \pm 0.2 \times 10^{-6}$ erg cm$^{-2}$ (15$-$150 keV), and a duration of $T_{90} = 225 \pm 10$ seconds \citep{sak05}. The energy spectral index was $0.34 \pm 0.06$ \citep{sak05}. The narrow field instruments XRT \citep{bur05} and UVOT \citep{mas05} observed the field of \object{GRB 050904} about 160 seconds after the trigger. No optical counterpart was observed within the UVOT field of view, while an X-ray transient was detected at position 00$^h$ 54$^m$ 50.4$^s$ +14$\degr$ 05$\arcmin$ 08.5$\arcsec$ \citep{cum05}. The first optical detection was made by TAROT \citep{boe99}, which observed a faint optical afterglow in the unfiltered frames \citep{klo05a, klo05b, boe05}. Other observations made with the BOOTES-1B telescope failed to detect this afterglow in the R band \citep{jel05}. Larger optical and infrared telescopes imaged the field of view 3 hours after the burst and detected an infrared counterpart \citep{hai05}. The absence of any counterpart observed from the R to U bands implied a large absorption, possibly due to the Lyman alpha forest \citep{hai05}. This was confirmed by the spectroscopic redshift measured by the Subaru telescope \citep[z = 6.29][]{kaw05}, which implied that the Lyman alpha cut-off was redshifted in the optical up to 8800 \AA. This burst is currently the more distant burst ever observed.

\section{Data reduction and analysis}
\label{sec_ana}

\subsection{X-ray data}
\subsubsection{Data reduction}

We obtained the XRT data from the SWIFT archive\footnote{see http://swift.gsfc.nasa.gov/docs/swift/archive/} and reduced them using the available packages and calibration files (SWIFT FTOOLS version 2.2 and CALDB version 20051028). Data were filtered using the provided good time intervals (sun and moon constraints, bright earth and South Atlantic Anomaly limitations) and standard criteria : CCD temperature $< -50\degr$C; grades 0 to 2 and energy above 0.3 keV for Window Timing mode (WT); grades 0 to 12 and energy above 0.5 keV for Photon Counting mode (PC). Note that the spectral ranges are different in WT and PC mode in order to take into account some uncertainties on the PC mode calibration \citep{osb05}. We extracted spectra and light curves using circle (PC mode) or box (WT mode) regions of 25 pixels radius (that enclose $\sim 95$ \% of the PSF, as explained in the XRT data reduction guide\footnote{available online at http://swift.gsfc.nasa.gov/docs/swift/analysis/}). The background was estimated using a larger region free of sources.

Our filtering criteria are mode dependent due to the change in the spectral range, and the mode switch may introduce some problems due to cross calibration uncertainties. Thus, any count light curve extracted need to be corrected for this dependency. To do so, we constructed a color-color diagram to look for spectral variability. This information allowed us to convert the count light curve into a flux one which is mode independent. This light curve is presented in Fig. \ref{fig_lc}. As one can see, the transition between the two modes occurs at the end of a large flare; however, the continuity of the light curve made us confident on our analysis.

We extracted spectra for several time intervals listed in Table \ref{table_x}. The spectra were re-binned to contain at least 20 net counts per bin using GRPPHA, and were fitted using XSPEC version 11.3.1 \citep{arn96}. All errors for the fit parameters quoted in the paper are given at the 90 \% confidence level for one interesting parameter.

\subsubsection{Data analysis}
\label{sec_spectra}

\begin{figure}
\centering
\includegraphics[width=9cm]{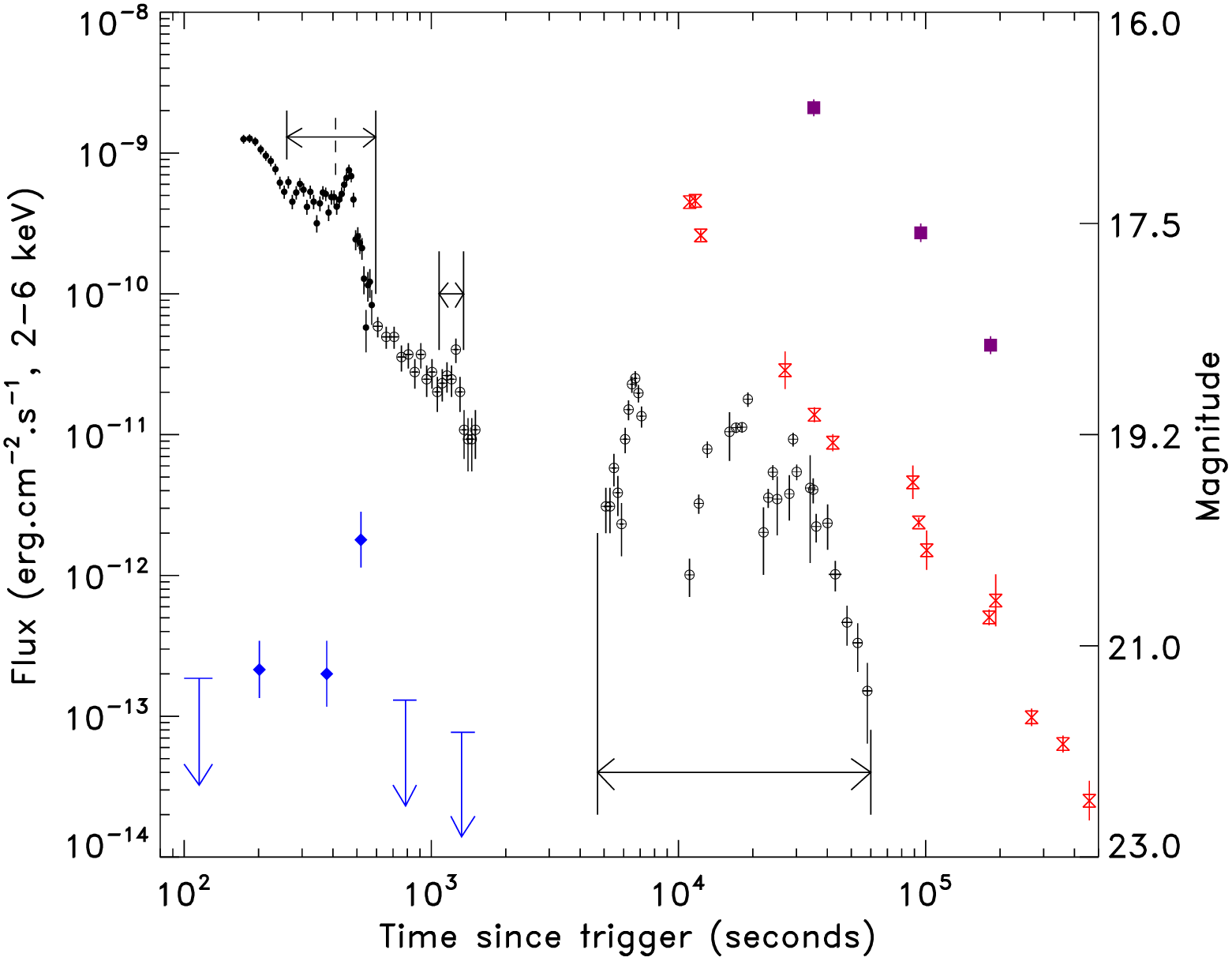}
\caption{Multi-wavelength light curve of \object{GRB 050904}. We present the X-ray data (black points) taken in the 2.0-6.0 keV band with filled circles (WT mode data) and open circles (PC mode data), the TAROT $I_T$ band data with blue diamonds, the $H$ data with solid purple squares \citep[from][]{hai05, tag05} and the $J$ data with open red symbols \citep[from][]{hai05, tag05}. For clarity, offsets by +6 and -1.5 magnitudes have been applied to TAROT data and $H$ band data respectively. We indicate on the figure the intervals excluded for the temporal fit of the X-ray data (solid horizontal arrows). The vertical dashed line indicate the position of the transition between the plateau and the first flare. See electronic version for colors.}
\label{fig_lc}

\end{figure}

The X-ray light curve shows a very complicated evolution, with several flares. In the following, we take as the start of a flare the point from which the derivative of the decay law becomes positive. The end of a flare is defined as the point where the light curve shows a flattening (we consider the flattening as the recovery of the continuum level). The early X-ray light curve ($t < 2000$ s) shows a steep decrease, followed by a plateau and a large flare (see Fig. \ref{fig_lc}). A second flare occurs at 1240 seconds \citep{man05}. We fitted this part of the light curve excluding all data points between the start and the end of all flares. Since the estimation of the start date of the first flare is complicated by the presence of the plateau, we conservatively excluded also the data located in the plateau (see Fig. \ref{fig_lc}). Since a sudden softening of the spectrum is observed at the time of the first flare, we have fit separately the data before and after the flare using a power law. The early decay index is $2.5 \pm 0.2$ ($\chi^2_\nu = 1.05$, 13 d.o.f.) and the late one $1.5 \pm 0.2$ ($\chi^2_\nu = 0.72$, 14 d.o.f.).

The whole spectra (0.3 keV or 0.5 keV to 10 keV) were fitted with a simple power law (i.e. $F(\nu) \propto \nu^{-\alpha}$), absorbed by our galaxy \citep[$N_H$ value fixed to the galactic one of $4.97\times 10^{20}$ cm$^{-2}$][]{dic90} and by local absorbers at redshift 6.3. We detect an excess of absorption before 264 s. After that, we have only upper limits, except between 384 and 463 seconds (which corresponds to the flare maximum). In some cases, the XRT may experience some gain variation for bright sources due to a bias subtraction problem \citep{cam06}. In such a case systematic deviations from the model appear near the oxygen edge (0.5-0.6 keV). Excluding the spectral bins between 0.5 and 0.6 keV from the fit of the flare spectrum, the $N_H$ is found to be $2.4^{+4.3}_{-2.4}\times 10^{22}$ cm$^{-2}$, i.e. compatible with 0.0. Note however that this effect does not influence the spectral index measurement.

We have tested if the hypothesis of a constant N$_H$ agrees with the data : we rejected it at the 99.5 \% confidence level using a $\chi^2$ test ($\chi^2$=13.41/3 d.o.f., using $1\sigma$ errors in the $\chi^2$ computation) and excluding from the fit the bin showing an increase of $N_H$ at t=420 s. Because of the large redshift, we cannot set a meaningful constraint on the host column density with the PC mode restricted to energies above 0.5 keV, as already noted by \citet{wat05}. Thus, in order to better constrain the spectral index, we fixed the extragalactic column density value to zero for the PC mode (after 596 seconds). Results from this spectral analysis are listed in Table \ref{table_x}. These results are in agreement with those published by \citet{wat05}. We note however that these authors reported an increase of the $N_H$ for $t> 9000$ seconds. An inspection of the residuals indicated a very poor fit between 2000 and 10000 seconds ($\chi^2_\nu = 2.04$, 42 d.o.f.). Adding some absorption locally to the burst fails to improve significantly the fit ($\chi^2_\nu = 2.03$, 41 d.o.f.), and resulted only in an upper limit \citep[$N_H < 1.2 \times 10^{23}$ cm$^{-2}$; however, this upper limit is consistent with the value reported by][]{wat05}. This poor fit could be due to some spectral component of the flares occurring between 2000 and 10000 seconds, thus we do not investigate it further.

\begin{figure}
   \centering
   \includegraphics[width=9cm]{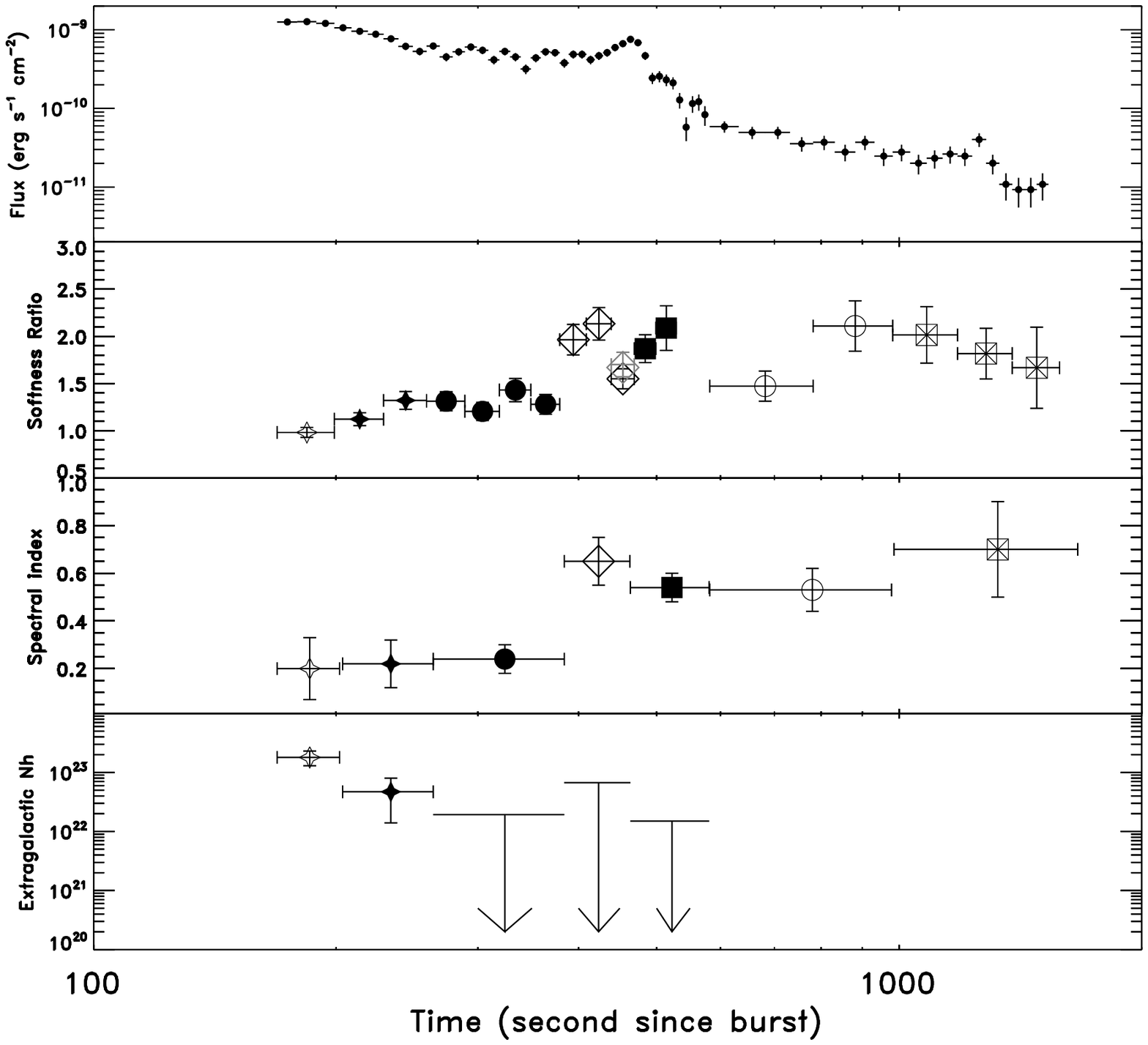}
   \caption{First panel (Top) : X-ray light curve of the early afterglow of \object{GRB 050904} in the 2.0-6.0 keV band. We have converted the count rate in flux using the conversion factors derived from our spectral analysis. Second panel) : The softness ratio of the X-ray afterglow (0.5-2.0 keV band versus the 2.0-10.0 keV band). We corrected the bin corresponding to the flare maximum (affected by an instrumental problem) by substituting the observed counts in the 0.5-0.6 keV band by the count rate predicted by the spectral model in the same band. The corrected value is indicated by the gray symbol. Third panel) : The energy index of the afterglow spectra. We assumed for each spectra an absorbed power law model. Last panel (Bottom) : the extragalactic column density obtained from the fit to the data. Note that the current calibration of the PC mode does not allow us to fit this quantity in the PC spectra. The symbols in panels b), c), and d) refer to the temporal slices listed in Table \ref{table_x}.}
              \label{fig_x}%
    \end{figure}

We restricted the spectral analysis to the early data ($t<$2000 s). Figure \ref{fig_x} shows the variations of the softness ratio, the spectral index and the extragalactic column density as a function of time. The softness ratio is computed by dividing the counts detected in the 0.5-2.0 keV band by those detected in the 2.0-10.0 keV band (i.e. the soft band is affected by absorption, while the hard band is not affected). We observe a global softening of the spectrum with time. Figure \ref{fig_x} also shows the variations of the energy spectral index, which is constant before the first flare, increases at the start of this flare, and then remains constant. A spectral fit on all the WT data before and after the start of the flare indicate a power law energy index of $0.23 \pm 0.05$ and $0.57 \pm 0.06$ respectively ($\chi^2_\nu = 1.2$, 288 d.o.f.). This latter value is consistent with the one derived from the following PC data ($0.53 \pm 0.09$).

\begin{table}
\caption{X-ray (0.3 or 0.5 keV to 10.0 keV band) spectral fitting results. The $\chi^2_\nu$ value of the fit is 1.26 for 463 degrees of freedom. The Extragalactic absorption is given at $z=6.3$.}             
\label{table_x}      
\centering                          
\begin{tabular}{ccc}        
\hline\hline                 
Temporal slice & Extragalactic & Energy index   \\    
               & absorption   &   $\alpha$      \\
   (s)         &($10^{22}$ cm$^{-2}$)&          \\
\hline                        
  169-202  & $18 \pm 5$          & $0.2 \pm 0.2$    \\
203-264    & $4.7^{+3.3}_{-2.8}$ & $0.2 \pm 0.1$   \\
264-384    & $<1.93$ ($0.1^{+1.83}_{-0.1}$) & $0.24 \pm 0.06$ \\
384-463    & $2.4^{+4.3}_{-2.4}$ & $0.65 \pm 0.1$  \\
464-581    & $<1.50$ ($0.0^{+1.50}$) & $0.54 \pm 0.06$ \\
596-979    & ---                 & $0.53 \pm 0.09$ \\
986-1666   & ---                 & $0.7 \pm 0.2$   \\
2000-10000 & ---                 & $0.73 \pm 0.08$ \\
10000-20000& ---                 & $0.74 \pm 0.07$ \\
$>20000$   & ---                 & $0.82 \pm 0.08$ \\
\hline                                   
\end{tabular}
\end{table}

\subsection{Optical data}
\label{sec_optical}

TAROT ({\it T\'elescope \`{a} Action Rapide pour les Objets Transitoires}) is a robotic observatory designed for GRB early detections \citep[e.g.][]{klo05}. 

The Gamma-ray burst Coordinate Network (GCN) distributed the SWIFT alert 81\,s after the beginning of the prompt emission of \object{GRB 050904}. The TAROT first exposure begun five seconds later. Twenty two unfiltered CCD images were taken between UTC 01:53:10.2 (t$_{trig}$+86.2s) and 02:08:09.6 (t$_{trig}$+1666.4s). Two technical problems occurred during the record: 
\begin{itemize}
\item[i)] A bug of the scheduling software lead to stop observations in the range t$_{trig}$ +254 to +312s. 
\item[ii)] The CCD camera cooler stopped due to a wire that broke during the fast slew of the telescope. As a consequence, the temperature of the CCD remained at $-50\degr$C until t$_{trig}$+260s and then increased continuously to +14$\degr$C at t$_{trig}$+1500s. Frames taken after t$_{trig}$+590s are strongly affected by thermal noise. 
\end{itemize}

We co-added frames into 6 final images in order to increase the signal to noise ratio.

\begin{table}[htb]
\caption{
Reference star characteristics nearby the position of GRB~050904.
}
\begin{center}
\begin{tabular}{c c c c c}
\hline\hline
2MASS & I & V-R & V-I & Sp. Type\cr
\hline
013.693725 & 11.68 & 0.70 & 1.31 & K4V \cr
013.732724 & 12.29 & 0.49 & 1.01 & K2V \cr
013.782866 & 12.39 & 0.35 & 0.74 & G0V \cr
\hline
\end{tabular}
\label{tarotref}
\end{center}
\end{table}

Calibrations of TAROT data are difficult because images were acquired with no filter, and the high $z$ of the GRB induces a non standard spectrum over the optical wavelengths. First, we choose three stars separated by less than 5 arcmins from the GRB and we calibrated their $VRI$ magnitudes with the 80\,cm telescope at the {\it Observatoire de Haute-Provence}, using stars of the SA92 field of calibration \citep{lon03}. We completed this study with spectra taken with the same telescope equipped by a low resolution spectrograph (R=150), and fit the star continua. Results of photometry, spectrometry and data from 2MASS catalog allowed us to attribute accurate magnitudes and spectral types (see Table~\ref{tarotref}). We computed the so called {\it superstar} as the sum of the three reference stars using the stellar spectral flux library provided by \citet{pic98}.
For each TAROT image, we computed the Point Spread Function (PSF) of the {\it superstar}. We fit the optical transient (OT) of \object{GRB 050904} by this PSF searching the $a$ coefficient corresponding to the minimum value of the (OT - $a \times$ PSF)$^2$.
To calibrate the flux of the OT, we assumed a spectral index $\beta$=-1.2 and a sharp cut--off at $\lambda$=8862 \AA. We modeled the response of the system (atmosphere, optics and CCD) and we applied it to the theoretical spectra of the OT and to the {\it superstar}. Then, we adjusted the flux of the OT at $\lambda$=9500 \AA~to obtain the same $a$ coefficient than that obtained from TAROT images (we call this ``filter'' the $I_T$ band). Results are reported in Table~\ref{tarotobs}, and presented together with the X-ray data in Fig. \ref{fig_lc}. As one can see, and as already reported in \citet{boe05}, TAROT has observed a flare in the $I_T$ band which is coincident with one X-ray flare.

\begin{table}[htb]
\caption{
Flux derived from TAROT images. In parenthesis, upper limits computed from PSF fit using a signal-to-noise (S/N) limit 3.
}
\begin{center}
\begin{tabular}{c c c}
\hline\hline
t$_{trig}$ range (s) & Flux at 9500\AA~(ergs cm$^{-2}$ s$^{-1}$ \AA$^{-1}$) & S/N\cr
\hline
  86 -  144 & (7.5$\times 10^{-16}$) $<$5.3$\times 10^{-15}$ & $<$2 \cr
 150 -  253 &  4.6 - 7.4 $\times 10^{-15}$ & 3.7  \cr
 312 -  443 &  4.2 - 7.4 $\times 10^{-15}$ & 3.1  \cr
 449 -  589 &  1.3 - 2.1 $\times 10^{-14}$ & 4.0  \cr
 595 -  978 & (4.2$\times 10^{-15}$) $<$4.2$\times 10^{-15}$ & 2.7  \cr
 985 - 1666 & (2.5$\times 10^{-15}$) $<$3.1$\times 10^{-15}$ & $<$2  \cr
\hline
\end{tabular}
\label{tarotobs}
\end{center}
\end{table}

We completed these optical observations with the ones reported in \citet{tag05} and \citet{hai05} (see Fig. \ref{fig_lc}), where a steep-flat-steep evolution was found. From an initial decay of $1.36 \pm 0.07$ between 0.125 and 0.5 days \citep{hai05}, the light curve flattens to a decay of $0.7 \pm 0.2$. At $2.6 \pm 1.0$ days, the light curve steepens to a decay of $2.4 \pm 0.4$, as usually observed in case of a so-called jet break \citep{rho97}.

\section{The spectral evolution of the X-ray afterglow}
\label{sec_nh}

The hard $\gamma$-ray emission is observed up to the start of the first X-ray flare \citep[t $\sim 420$ s,][]{sak05}. Moreover, the X-ray emission before the first flare (t = 169$-$400 s) has temporal ($\delta=2.45\pm0.2$) and spectral ($\alpha=0.2\pm0.1$) indexes compatible with the tail of the prompt one, i.e. with off-axis emission \citep[closure relationship $\delta=2+\alpha$,][]{kum00}, and the X-ray spectral index observed before the start of the flare is comparable to the gamma-ray one ($0.2\pm0.1$ vs  $0.34 \pm 0.06$ respectively). Thus, this part of the observations may be linked to the prompt emission. During this phase (t$< \sim 400$s), a significant progressive softening is observed, correlated with a decrease of the $N_H$ value. In fact, the spectral index does not vary during that time (see Fig. \ref{fig_x}). One may thus explain the softening as the consequence of the observed decrease of the column density, rather than the consequence of the well known hard-to-soft evolution of the prompt emission. Such a decrease has been already observed in previous bursts, such as \object{GRB 050730} \citep{sta05}, \object{GRB 000528} \citep{fro04} and \object{GRB 980329} \citep{fro00, laz02} although, in the latter two cases, the available statistics were too low to exclude a constant $N_H$. The decrease of the measured equivalent column density that we observe in the case of GRB 050904 may be the evidence of a progressive photo-ionization of the initially cold gas in which the GRB occurs by the burst itself \citep{per98}.

A significant softening is observed at the start of the first flare; the softness ratio remains constant throughout the flare except at its maximum ($\sim460$ s) where its value changes from $1.95$ to $1.55\pm0.1$ ($1\sigma$ error). This latter variation may be due to a calibration problem (see Sec. \ref{sec_spectra}). Correcting for this effect (see Fig. \ref{fig_x}), we obtain a softness ratio of $1.67\pm0.14$ ($1\sigma$ error), consistent with no spectral variation.

\begin{figure}
   \centering
   \includegraphics[width=9cm]{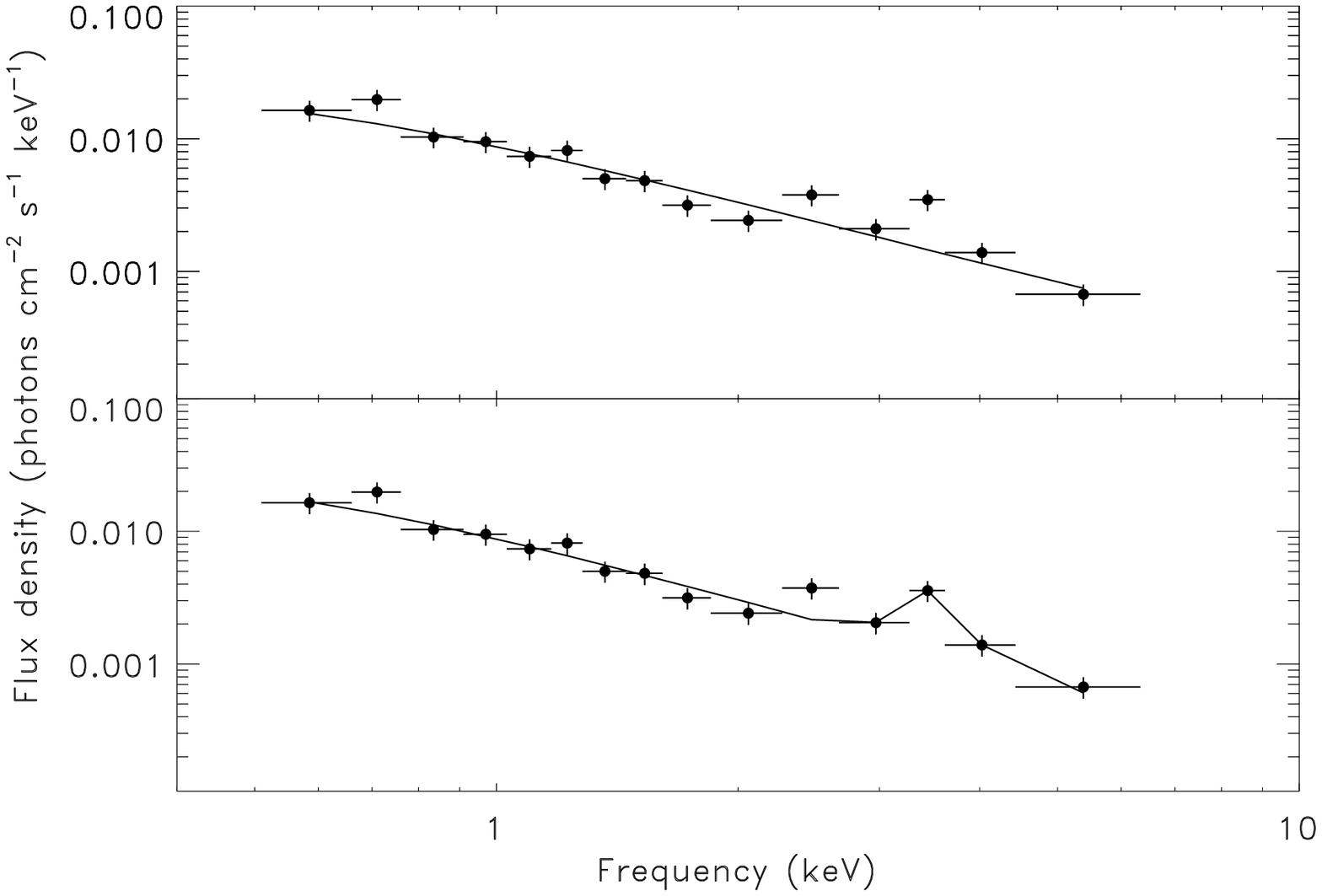}
   \caption{The X-ray spectrum of \object{GRB 050904} integrated between 582 and 978 seconds after the burst. Top : fit with a single power law. Bottom : fit with a power law continuum and a Gaussian line.}
              \label{figure_spectre}%
    \end{figure}

At t=582 s, another hardening is observed. Very surprisingly, the spectral index does not vary and no excess of absorption is detected at that time. However, the spectral fit of this part of the data is very poor ($\chi_\nu^2=2.1$, 13 d.o.f.). A fit with a two component model (e.g. a hard power law plus a soft power law, as expected if we observe the afterglow together with a significant signal from the tail of the prompt emission) is even worse ($\chi_\nu^2=2.4$, 11 d.o.f.). As can be seen from the top of Fig. \ref{figure_spectre}, the main discrepancy arises from a bin at about 3.3 keV. We have tried to take this discrepancy into account by adding a narrow line to our spectral model rather than simply ignoring the spectral bin. We obtain a better agreement with the line parameters  $E = 3.46 \pm 0.15$ keV and $\sigma < 0.46 $ keV ($\chi^2_\nu = 1.34$, 10 d.o.f.). The continuum parameter changes to $\alpha = 0.66 \pm 0.17$. Using this continuum model, we match a softness ratio of 1.9, compatible with the hypothesis that the softness ratio is constant after the first flare. The observed change of the softness ratio at t=582 s in Fig. \ref{fig_x} may thus be related only to that discrepancy at high energy. The statistic does not allow us to discriminate between a spurious deviation and a real emission line. This deviation could be due to instrumental problems (such as background fluctuations) rather than a real physical process.

After the initial 2 ks of data, the XRT light curve features several flares (see Fig. \ref{fig_lc}). X-ray spectra extracted during the 3rd, 4th and 5th flares indicate a small softening of the spectral properties. We cannot investigate the underlying decay law due to the large flaring activity.

\section{The stellar progenitor and the surrounding medium of \object{GRB 050904} : observation of a termination shock}
\label{sec_discu}

We have used the closure relationships given by \citet{che00}, \citet{Sar98}, \citet{Sar99} to investigate the surrounding medium of \object{GRB 050904} and on the fireball geometry. We considered the X-ray data between the end of the first flare (t$\sim$580 s) and $\sim 1666$ s. In the optical band, we separate the data between {\it early} and {\it late} ones, as observed between 0.1 and 0.5 days, and between 0.5 and 2.6 days respectively. The transition date (0.5 days) is chosen because both \citet{hai05} and \citet{tag05} observe a flattening at that time. The {\it early} optical data occur during the 4th and 5th X-ray flares. We thus prefer to conservatively discard these data, and use only the {\it late} ones for the closure relationship study, adopting the $\alpha$ and $\delta$ values given by \citet{tag05}. We list the closure relationship results in Table \ref{table_closure}.

The X-ray data are marginally consistent with the jet hypothesis. However, in this case, one should not expect any further steepening in the light curve \citep{rho97}, contrary to that observed by \citet{tag05} at about $2.6$~days.

\begin{table*}
\caption{Closure relationships in the standard fireball model computed using the spectral and temporal information. We give the relationships for the $J$ band after 0.5 days. The specific frequency is $\nu_c$ (the cooling frequency) in the slow cooling regime, or $\nu_m$ (the injection frequency) in the case of fast cooling. All errors are quoted at the 90 \% confidence level.} 
\label{table_closure}      
\centering                          
\begin{tabular}{ccccccc}        
\hline\hline                 
Medium class    & Cooling & Specific      & Closure               & Expected &   X-ray        &  late \\
and             & regime  & frequency     & relationship          & value    &                &  optical \\
geometry        &         & position      &                       &          & (0.5-10.0 keV) &  ($J$ band) \\
                &         &               &                       &          & (582-1666 s)  &  (0.5-2.6 days) \\
\hline
Isotropic Wind  & Fast    & $\nu_m<\nu$ & $\delta - 1.5 \alpha$ &  $-0.5$  & $0.6\pm0.3$    &  $-1.1\pm0.6$ \\
                &         & $\nu_m>\nu$ & $\delta - 0.5 \alpha$ &  0.0     & $1.2\pm0.3$    &  $0.1 \pm0.4$ \\
                & Slow    & $\nu_c<\nu$ & $\delta - 1.5 \alpha$ & $-0.5$   & $0.6\pm0.3$    &  $-1.1\pm0.6$ \\
                &         & $\nu_c>\nu$ & $\delta - 1.5 \alpha$ &    0.5   & $0.6\pm0.3$    &  $-1.1\pm0.6$ \\
Isotropic ISM   & Fast    & $\nu_m<\nu$ & $\delta - 1.5 \alpha$ & $-0.5$   & $0.6\pm0.3$    &  $-1.1\pm0.6$ \\
                &         & $\nu_m>\nu$ & $\delta - 0.5 \alpha$ & 0.0      & $1.2\pm0.3$    &  $0.1 \pm0.4$ \\
                & Slow    & $\nu_c<\nu$ & $\delta - 1.5 \alpha$ &  $-0.5$  & $0.6\pm0.3$    &  $-1.1\pm0.6$ \\
                &         & $\nu_c>\nu$ & $\delta - 1.5 \alpha$ &  0.0     & $0.6\pm0.3$    &  $-1.1\pm0.6$ \\
Jetted fireball & Slow    & $\nu_c<\nu$ & $\delta - 2 \alpha$   &  0.0     & $0.4\pm0.3$    &  $-1.7\pm0.8$ \\
                &         & $\nu_c>\nu$ & $\delta - 2 \alpha$   &  1.0     & $0.4\pm0.3$    &  $-1.7\pm0.8$ \\
\hline
\end{tabular}
\end{table*}

\begin{figure*}
   \centering
   \includegraphics[width=13cm, angle=-90]{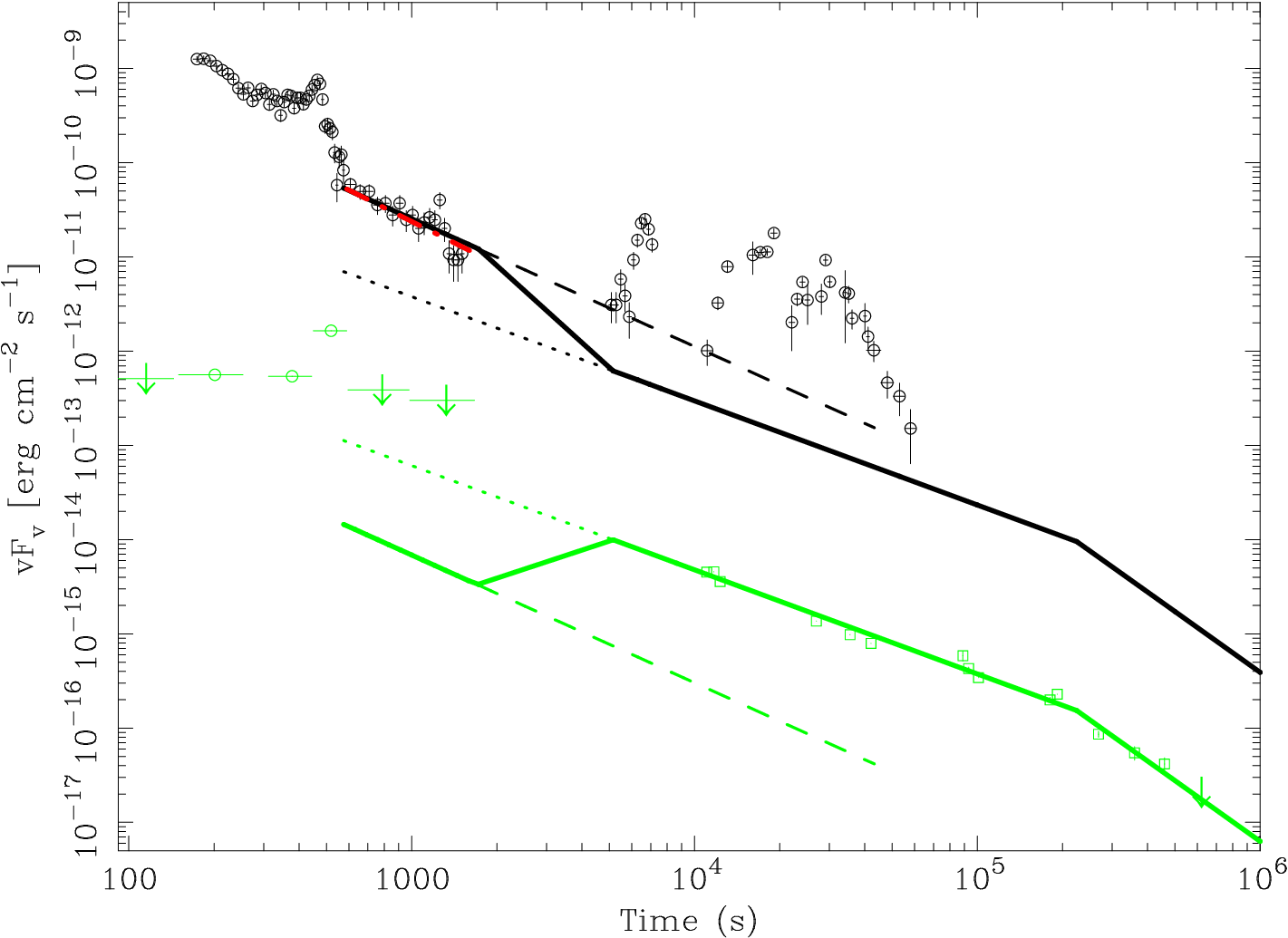}
   \caption{Global modeling of the broad-band data using a wind ($A_*$ = 1.8, $\epsilon_B$ = $10^{-6}$, $\epsilon_e$ = $8.5\times10^{-2}$) plus ISM ($n$ = 680, $\epsilon_B$ = $0.015$, $\epsilon_e$ = $10^{-2}$) scenario (solid lines). We present in black the X-ray data and in green the optical data \citep[circles are TAROT data rescaled in the J band, squares are data taken from ][rescaled by a factor 0.01 for clarity purpose]{tag05,hai05}. The red-dot-dashed line is the best fit decay law to the X-ray data between $\sim 600$~s and $\sim 1700$~s. The dotted lines represent the extrapolation of the ISM model by \citet{fra06} to early times. The dashed lines are the extrapolation of the wind model to late times. We assumed an instantaneous wind-to-ISM transition at $t\sim 1700$~s (see text), smoothed by the effect of the fireball curvature ($\Delta t\sim 3400$~s). See electronic version for colors.}
              \label{fig_frail}%
    \end{figure*}

An ISM scenario (slow cooling phase with $\nu_{c}$ above the X-ray band) is only in marginal agreement with the early X-ray data (t$<$2000s, see Table \ref{table_closure}). Moreover, the expected difference between the X-ray and optical light curve decay indexes should be $\delta_{x}-\delta_{o}=0$ or $\delta_{x}-\delta_{o}=0.25$. The observed value is $\delta_{x}-\delta_{o}=0.8\pm0.4$, again only marginally consistent with the expectations. On the other hand, the late broad-band (radio-to-X-rays) observations clearly agree with an ISM environment having a very high density and $\nu_c$ below the optical band, as recently shown by \citet{fra06}. This model predicts that the cooling frequency is below the optical band even at early ($\sim$ 500) times. However, its extrapolation in the X-ray band at these times (black-dotted line in Fig. \ref{fig_frail}) cannot describe the data: the predicted flux is at least one order of magnitude below the observed one and the extrapolated light curve is flatter than the observed best fit decay (red-dash-dotted line), as already anticipated above. Using the closure relationships, the hypothesis that the interstellar medium explains the X-ray data is excluded at the 5.8$\sigma$ level. We thus conclude that while an ISM describes well the late broad-band observations \citep{fra06}, it does not reproduce satisfactorily the early X-ray data (t$<$2000s).

A wind environment is clearly favored by the early (t$<$2000s) X-ray observations ({\it slow cooling, $\nu_{c}$ above the X-ray band}, see Table \ref{table_closure}). In a wind medium $\nu_{c}$ increases with time \citep{che00}, thus we expect also the optical band to be below this frequency at late times. However, this hypothesis is rejected at the 4.2$\sigma$ level. A similar case with a flat spectral index and a typical temporal decay was observed only for \object{GRB 040106}, and attributed to a wind density profile \citep{gen05a}. Combined with the results by \citet{fra06}, this suggests that the fireball could be expanding into a medium whose density profile at small distances is that expected for a wind ($n\propto r^{-2}$), and becomes constant at larger radii. 

In this hypothesis, a wind termination shock is expected to mark the transition between the two different environments. The termination shock radius, $R_{t}$, is defined as a function of the wind density parameter $A_*$ and the ISM density $n$ \citep{che04,pan04}: 

\begin{equation}
R_t = 1.1 \times 10^{18} A_{*}^{0.5} n^{-0.5}
\label{eq:AleeAle1}
\end{equation}

where $R_{t}$ is in units of $cm$. The termination shock crossing time can be directly estimated from the observations to be located between 0.019 days (the latest time at which the X-ray closure relationships can be computed and favor a wind environment) and 0.5 days (the earliest time at which the optical data favor an ISM). One can then use the standard fireball prescriptions that give the fireball radius as a function of time, initial energy $E_{52}$ (in the remaining we will use the standard notation $A_x = 10^x A$) and $A_*$ to derive an independent estimate of the termination shock radius as a function of these two quantities \citep{che04}:

\begin{equation}
R_{t}=1.55\times10^{17} E_{52}^{0.5} A_{*}^{-0.5} \left(\frac{t_{*,d}}{1+z}\right)^{0.5}
\label{eq:AleeAle2}
\end{equation}

where $t_{*,d}$ is the shock crossing time in units of days. Combining equations (\ref{eq:AleeAle1}) and (\ref{eq:AleeAle2}), one can derive the possible values of $A_*$:

\begin{equation}
\label{eq_A*}
 E^{0.5}_{52} n^{0.5} 7.24 \times 10^{-3} < A_* < E^{0.5}_{52} n^{0.5} 3.69 \times 10^{-2} 
\end{equation}

Using the best fit results by \citet{fra06}, $n$=680 and $E_{52}$=88, we obtain $1.77 < A_* < 9.03$.

We thus derive the position of the termination shock radius, $R_t$, to be $0.018~\mathrm{pc} < R_t < 0.041~\mathrm{pc}$. Note that this burst is very energetic \citep{kan06} and surrounded by a very dense medium \citep{fra06}, and thus the termination shock position of \object{GRB 050904} may not be typical. However, this value is compatible with the upper limits obtained by \citet{che04} on several bursts.

Using the prescriptions given by \citet{yos03}, we have checked if there is a solution compatible with the X-ray data within the range of values for $A_*$ derived using the best fit by \citet{fra06}. As shown in Fig. \ref{fig_frail}, setting $A_{*}=1.8$, we nicely describe the early X-ray data (black solid line before 2000 s). This implies a termination shock radius of $R_t \sim 1.8 \times 10^{-2}$ pc. Note that we find $\epsilon_B$ = $10^{-6}$ and $\epsilon_e$ = $8.5\times10^{-2}$, that are different from Frail's values. However, such a change is reasonable because of the shock presence. In fact, the microphysic parameters are allowed to vary during the fireball evolution even if there is no termination shock \citep{pan06}. Thus, a wind plus ISM scenario can provide a global model for the broad-band observations. 

With these parameters, we find a wind density of $\sim 170$ cm$^{-3}$ at the termination shock radius, giving a density jump of $\sim 4$ as implied by eq. \ref{eq:AleeAle1}.

Finally, an order of magnitude estimate of the intrinsic absorption implied by the wind environment is given by \citet{piro05} :

\begin{equation}
N_{H,22}=3\frac{A_*}{r_{13}}
\label{eq_fun}
\end{equation}

where $r_{13}$ is the radius at which the prompt emission is produced. Using $A_*=1.7$ and assuming $r_{13} = 1$, we find consistency with the value we observed at the start of the X-ray observation.

\section{Multi-wavelength modeling of the first flare}

With the Swift observations, it appears clear that flares are very common in GRB light curves during a time interval that goes from hundred to thousand of seconds \citep{obr06}. The evidence of such a common flaring activity has favored the development of several models that attempt to explain the nature of X--ray flares, both in the framework of Internal Shock and External Shock. In the following, we comment on the first flare observed simultaneously in the X-ray and optical bands in light of these models.

\subsection{Delayed external shock scenario}
\label{sec_galli}

The X-ray spectrum of the first flare is softer than the preceding plateau and is consistent with the following afterglow emission. Similar properties have been previously observed in other bursts \citep[e.g \object{GRB 011121}, \object{XRR 011211} and \object{XRF 011030},][]{piro05, gal05}, and have been explained in the framework of a delayed external shock, expected e.g. in a thick shell fireball scenario \citep{Sar99}. In this framework both the X-ray and the optical flares are produced by the external shock, thus explaining the coincidence between the two flares. 

Very recently, \citet{laz05} have shown that a central engine releasing most of its energy reservoir during the final stages of its activity ($t_{eng}$), implies that the afterglow emission is described by a simple power law only if the origin of the time $t_0$ is of the order of $t_{eng}$. This implies that the beginning of the afterglow emission $t_0$ is the time of flare appearance $t_{flare}$ \citep{gal05}. We used the model of \citet{gal05}, that take this effect into account, to check if it can explain the multi-wavelength flare.

\begin{figure}
\centering
\includegraphics[width=6.3cm,angle=-90]{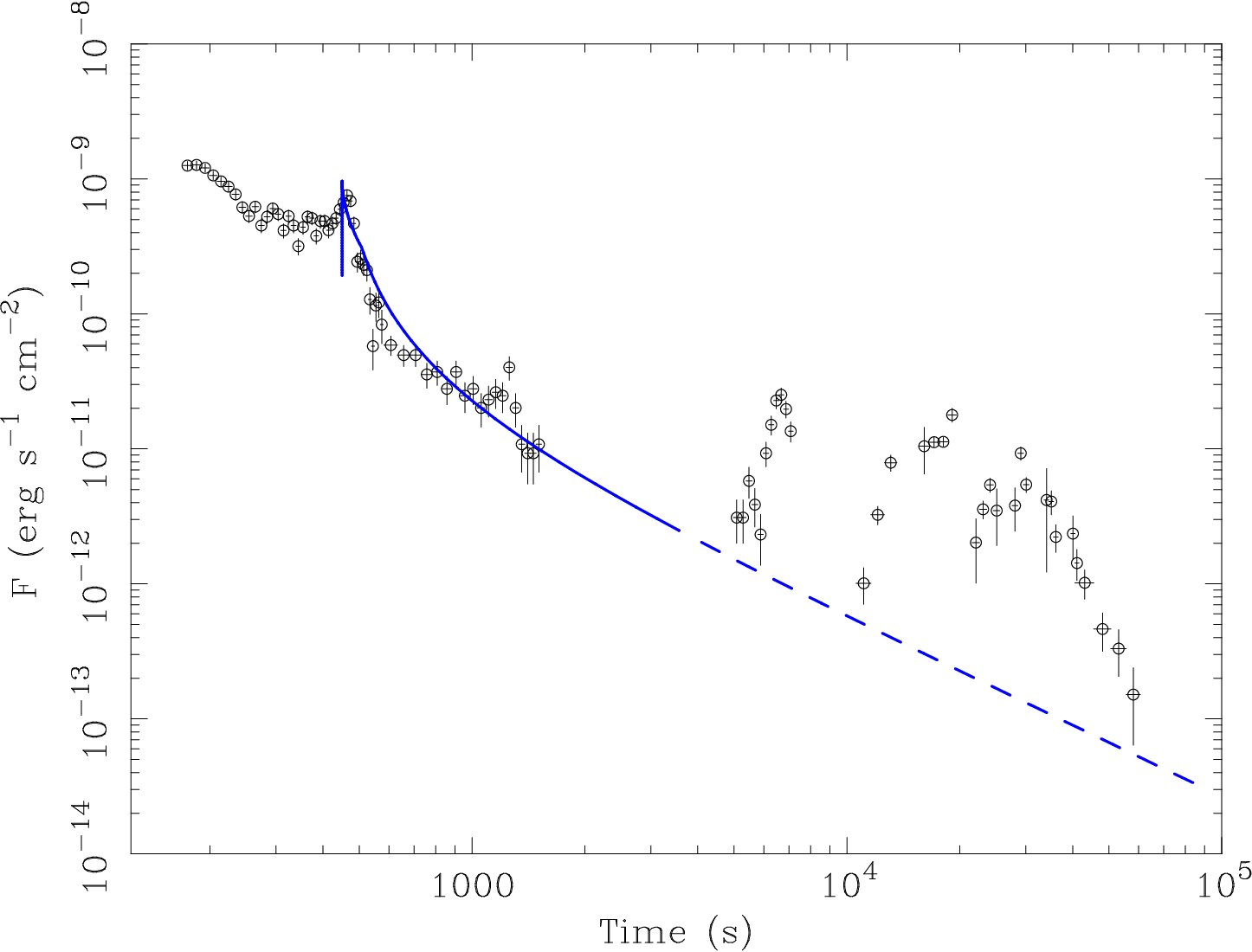}
\caption{X--ray light curve of \object{GRB 050904} produced by a thick shell
fireball expanding in a wind with the onset of the afterglow shifted to $t_0=464$ s. 
The model parameters are $E_{52}$=200, $\Gamma_0$=200, $A_*$=0.5,
$\epsilon_e$=0.008, $\epsilon_B=10^{-6}$, $p$=2.1.}
\label{wind_x_relchius}
\end{figure}

This model is a single component model that does not take into account the contribution of the prompt emission (that could affect the rise of the flare), nor the contribution from the reverse shock (that does not cross instantaneously the thick shell). Thus, it can fail to describe the rising of the flare emission, but it can be applied to the decay of the first flare and the following X-ray data (excluding all subsequent flares). In the case of a fireball expanding in a wind with $\nu < \nu_c$, $\alpha=(p-1)/2$ \citep{che04}. We derive $\alpha \sim 0.55$, which implies $p=2.1$. We have fixed  $E_{52}$=200. The remaining model parameters, i.e. $A_*$, $\epsilon_e$ and $\epsilon_B$ are determined in order to describe the data. We find a solution with parameters $\Gamma_0$=200 (where $\Gamma_0$ represent the initial Lorentz factor), $\epsilon_e$=0.008, $\epsilon_B=10^{-6}$ and $A_*=0.5$ (see Fig. \ref{wind_x_relchius}). This value of $A_*$ implies $1 < n < 20$, compatible with the range $ 10 < n < 10^4$ given by \citet{fra06}. 

However, some inconsistencies are implied by this scenario. First of all, in a thick shell, the afterglow peak is at the time when the prompt emission ceases \citep{Sar99}. As discussed in Sect. \ref{sec_nh}, the data preceding the flare are consistent with high latitude emission observed after the end of the prompt phase. Under this assumption, there is a temporal gap between the end of the prompt emission and the flare. Thus, such a gap is not compatible with relating the appearance of the flare with the onset of the afterglow from a thick shell.   

Moreover, even without considering the data before the flare, no solution can reproduce the optical observations in a wind case : the calculated optical flux at the time of the flare is $\sim 10^{-26}$ erg cm$^{-2}$ s$^{-1}$ Hz$^{-1}$, about 1.5 order of magnitude below the measured flux. In fact, with the cooling frequency $\nu_c$ above the X-ray band, the predicted broad band optical to X-ray spectral index value ($\alpha_{ox}$) is $\sim 0.6$. Instead, connecting the mean flux at 5 keV and the optical flux at the time of the flare, $\alpha_{ox}$ is $\sim 1.1$. As a consequence the extrapolation of the X-ray flux in the $I_T$ band falls below the optical data. A value of $\alpha_{ox} \sim 1.1$ at the time of the flare can be obtained assuming that $\nu_c$ is below the optical range during the flare, with $p=2.1$, but this is not compatible with the closure relationships.

\subsection{Possible two-component scenarios}
\label{sec_corsi}

\subsubsection{The role of the reverse shock}

\object{GRB 050904} is one of the few GRBs (the others are \object{GRB 990123}, \object{GRB 041219A}, \object{GRB 050401}, \object{GRB 060111B}, \object{GRB 060124}, \object{GRB 060904B}) for which a prompt optical emission was observed simultaneously with the high energy one \citep{Aker99,Fan05,Ryk05,boe05,Klo06,rom06, klo06b, uga06}. \citet{boe05} underlined the similarities between \object{GRB 990123} and \object{GRB 050904} in terms of optical flare brightness (once accounting for the difference in the distance), and broad-band spectrum of the prompt emission : in both cases the extrapolation of the high energy data to the optical band falls well below the optical points \citep{Cor05,boe05}, suggesting a different origin for the low and high energy emission mechanisms. In the case of \object{GRB 990123}, this hypothesis was further supported by the lack of coincidence between the peak observed in the optical and those observed in the high energy light curve \citep{Cor05}.

In the standard fireball model \citep{ree92, mes97, pan98}, when a relativistic ejecta moves into the cold ISM, two shocks form, an outgoing one that propagates into the ISM (the Forward Shock, FS) and a Reverse Shock (RS) that propagates into the ejecta \citep{Sar99}.

The typical synchrotron frequency of the RS emission is in the IR-to-optical region. If the signal observed in the low-energy part of the prompt spectrum is related to the RS, while prompt $\gamma$- and X-ray emission to internal shocks (or late internal shocks for the late time X-ray flares), one should not expect the low energy emission to fall on the extrapolation of the high energy one; moreover, there should not be any correlation between the pulses observed in the high energy light curve and the peak time of the optical flare. This is in fact what was observed for \object{GRB 990123}. In the case of \object{GRB 050904}, the high energy spectrum cannot be extrapolated at low energies, but the optical flare observed by TAROT appears to peak simultaneously with the first X-ray flare. In the internal shock-RS scenario, such a coincidence should be fortuitous. 

Recently \citet{Fan05} explored the possibility of relating X-ray flares to RS synchrotron emission appearing above the FS one, when the physical parameters ($\epsilon_{e}$ and $\epsilon_{B}$) in the RS and the FS are different. In this framework, one could think of relating both the optical and the first X-ray flare to RS emission. The temporal coincidence between those two flares would be a natural expectation of this model. However, \citet{Wei06} noted that even in this scenario, the X-ray spectrum should be a power-law extension of the optical emission, but the observations show that the optical-to-X-ray spectrum cannot be described by a simple synchrotron spectrum, even taking into account the possible presence of a spectral break between the optical and the X-ray band. 

\subsubsection{Flares in a RS synchrotron plus inverse Compton scenario}

Recently, \citet{Kob05} suggested the possibility of explaining X-ray flares via synchrotron self-Inverse Compton (IC) radiation from the RS. The prompt optical flare should be associated with synchrotron emission from the RS while the simultaneous X-ray flare should be the result of synchrotron photons being up-scattered via the IC process in the X-rays. The RS evolution depends on the initial Lorentz factor of the shell itself \citep{Sar97}. In the so-called thin shell case, the RS remains Newtonian during all the crossing time and the effective energy extraction takes place at the deceleration time $t_{d}$. On the other hand, if the initial Lorentz factor of the shell is sufficiently high, i.e. greater than the critical value $\Gamma_c$:
\begin{equation}
\label{eq_corsi}
\Gamma_{c}\sim 130 \left(\frac{1+z}{2}\right)^{3/8}E^{1/8}_{52}T^{-3/8}_{2}n^{-1/8}
\end{equation}
then the RS becomes relativistic while crossing the shell \citep{Kob05}. In Eq. \ref{eq_corsi},  $T_2$ is the burst duration. This corresponds to the thick shell case. In a thin shell fireball expanding in an ISM, both the low and high energy emissions should thus peak at $t_{d}$ and the temporal coincidence between the optical and the X-ray flares would be a natural expectation for this model too. In particular, in the thin shell case, the peak time of the X-ray and optical flares ($\sim 460$~s) constrains the model parameters according to the following relationship \citep{Kob05}:

\begin{equation}
t_{d}=190 \left(\frac{\Gamma_{0}}{80}\right)^{-8/3} \left(\frac{1+z}{2}\right)E^{1/3}_{52}\left(\frac{n}{5}\right)^{-1/3}~s \sim 460~s
\end{equation}

Setting that $E_{52}\sim 200$, $z\sim6.3$, for $n\sim 100$ it is $\Gamma_{0}\sim 120$ and $\Gamma_{c}\sim170$ for $T\sim 225$ s, consistent with the hypothesis of being in a thin shell case (i.e. $\Gamma_{0}<\Gamma_{c}$).

Using an additional emission mechanism like IC gives a natural explanation for the impossibility of extrapolating the spectrum of the X-ray flare down to the optical data points. In particular, we know that the $0.5-10$~keV spectrum of the flare is rather flat, with a spectral index of $\alpha\sim0.6$. Setting $p\sim 2.1$, the spectrum of the flare can be explained assuming that the X-ray band is between $\nu^{IC}_{m}=2\gamma^{2}_{m}\nu_{m}$ and $\nu^{IC}_{c}=2\gamma^{2}_{c}\nu_{c}$ around the deceleration time \citep{sar01}. Considering that $\gamma_{m}\sim 100$, if $\nu^{IC}_{m}\sim 1.0$~keV and $\nu^{IC}_{c}\sim 10$~keV, then $\nu_{m}\sim 10^{13}$~Hz and $\nu_{c}\sim 3 \times10^{13}$~Hz, that are reasonable values for the RS synchrotron break frequencies at the deceleration time \citep{Kob05}. With $p=2.1$, the RS synchrotron flux at $\sim 1$ keV is $\sim 1/1060$ times the optical one. The ratio between the peak flux observed by TAROT between $449$~s and $589$~s, and  the {\it mean flux} observed in the same temporal bin at $1$~keV is $\sim 1/750$ (i.e. a factor of $\sim 1.4$ greater than that extrapolated from the RS synchrotron emission), while the ratio between the same optical flux and the {\it peak flux} of the X-ray flare at $1$~keV is $\sim 1/280$ (i.e. a factor of $\sim 4$ greater than that extrapolated from the RS synchrotron emission). Thus, to explain the broad-band spectrum, the RS IC peak flux should be a factor of $\sim 1.4-4$ greater than the RS synchrotron flux at the same frequency (which is also compatible with the observed $2.0-6.0$ keV flux increase by a factor of $\sim 2$ in the rising part of the flare). As shown by \citet{Kob05}, an IC bump as high as 6 times the synchrotron RS emission can be explained with a reasonable choice of parameters. 

The model proposed by \citet{Kob05} thus appears to be a viable one to explain the broad-band spectrum and the temporal coincidence between the optical and the X-ray flares. However, one problem arises : explaining the first X-ray flare via IC emission with $\nu^{IC}_{m}<\nu_{X}<\nu^{IC}_{c}$, implies that the decaying part of the flare should be as $(t/t_{d})^{-(3p+1)/3}$ \citep{Kob05}, which for $p=2.1$ gives $\delta=2.4$, too shallow to agree with the observations (the X-ray emissions declines as $t^{-12.6}$ between 464 s and 534 s).

Moreover, in order for the RS emission to be dominant on the FS one, the peak frequency of the FS at the deceleration time should be $\nu^{peak}_{syn}<<\nu_{X}$; combining this with the fact that the spectral index of the early X-ray afterglow is rather flat, it should be min($\nu_{m},\nu_{c}$) $ <\nu_{X}<$ max($\nu_{m},\nu_{c}$) until $\sim10^{5}$ s. However, as shown in Table \ref{table_closure}, in this regime the expected temporal decay is too flat to be compatible with the observations.

Finally, the presence of a plateau before the optical flare observed by TAROT should also be considered: \citet{Wei06} shown that a plateau plus an optical flare could be produced if the outflow had a more complex structure (e.g. two components with different Lorentz factors, widths and isotropic energies). However, a fit to the optical light curve gives a value for $\epsilon_{B}$ too high for being compatible with the idea of explaining the X-ray flare as IC emission from the RS. In fact, a high magnetization suppresses the importance of IC emission with respect to synchrotron one. 

\subsubsection{Internal shock scenario}

\citet{Wei06} have recently shown how the problem of the steep temporal decay of the X-ray flare can be solved invoking a late internal shock model, where the optical flash comes from late internal shock synchrotron emission while the first X-ray flare is produced by late internal shock IC emission. In this model the temporal coincidence between the optical and the X-ray flare is a natural expectation as it is in the model proposed by \citet{Kob05} and there is no optical to X-ray extrapolation problem.

Finally, \citet{Zou05} interpreted the early-to-late time multi-band observations assuming that all the highly variable X-ray emission of \object{GRB 050904} originated from internal shocks developing when faster shells, continuously ejected by the central engine, overtake a slower one. In this model, the main contribution to X-ray emission comes from relativistic RSs while the optical emission is ascribed to Newtonian FSs. This model is very different from the standard assumptions of the canonical fireball model, but seems to be able to fit the prolonged flaring activity observed in the X-rays. However, we notice that the optical flare and the overall decay behavior of the X-ray light curve still need to be modeled in detail. We also underline that this model assumes a super-long central engine activity, i.e. that the shell ejection process lasts $10^{4}-10^{5}$~s.

\section{Conclusion}
\label{sec_conclu}

We have analyzed multi-wavelength observations of \object{GRB 050904} obtained with TAROT and SWIFT. Before the first X-ray flare, the X-ray decay and spectral indexes are consistent with the hypothesis of observing the tail of the prompt emission by curvature effect. After the flare, the X-ray emission is consistent with the hypothesis that the fireball expands in a wind medium. This would privilege a stellar progenitor. We estimate the wind density parameter to be $1.77 \lesssim A_* \lesssim 9.03$. After 0.5 days, the data are consistent with the hypothesis that the fireball is expanding in an interstellar medium. The hypothesis that a single kind of surrounding medium can explain all data is clearly rejected : the probability that the wind medium (which explains the X-ray data) can explain the optical data is 1.7$\times 10^{-5}$ (exclusion at 4.2$\sigma$) and the  probability that the interstellar medium (which explains the optical data) can explain the X-ray data is 4$\times 10^{-9}$ (exclusion at 5.8$\sigma$). This implies that the fireball may have crossed a termination shock. From the temporal constraints on the crossing time, we estimate the termination shock position to be $ 1.8 \times 10^{-2}$~pc~$\lesssim R_t \lesssim 4.1 \times 10^{-2}$ pc. We observe a significant excess of absorption at the start of the X-ray observation, $N_{H}=(18\pm5)\times 10^{22}$~cm$^{-2}$, compatible with the wind expectations. We further observed a decrease of this column density during the first seconds of the observation, significant at the $\sim 3\sigma$ level, that could be explained by photo-ionization of the medium by the burst. We investigated the simultaneity of the X-ray and optical flares in the context of several models. None of them can explain all the available data. In particular, the optical-to-X-ray spectrum cannot be described in a simple synchrotron scenario. Adding the contribution of IC emission in the X-ray band, a RS model fails to explain the steepness of the flare decay, while a late internal shock is favored.

Due to the time dilation associated with its large distance, \object{GRB 050904} offered for the first time the possibility of well sampling the prompt-to-afterglow transition, and the very early afterglow phase at several wavelengths. Its observation opened more questions than confirming canonical models. SWIFT is well suited to detect high redshift events, and hopefully other observations will be available to close those questions.

\begin{acknowledgements}
B.G. \& G.S. acknowledge supports from the RTN "GRB : an enigma and a tool". We acknowledge financial supports from the French GDR PCHE. The TAROT telescope has been funded by the CNRS, the INSU, and the Carlsberg Fundation. It has been built with the support of the {\it division technique} of the INSU. We thank the technical staff contributing to the TAROT project : G. Bucholtz, J. Eysseric, C. Pollas, Y. Richaud. We also thank G. Cusumano and O. Godet with help during the XRT data analysis.

\end{acknowledgements}

\end{document}